\documentstyle[12pt]{article}

\newlength{\dinwidth}
\newlength{\dinmargin}
\begin{document}
\newcommand{\be}{\begin{equation}}
\newcommand{\ee}{\end{equation}}
\newcommand{\bea}{\begin{eqnarray}}
\newcommand{\eea}{\end{eqnarray}}
\newcommand{\nn}{\nonumber}
\newcommand{\dd}{\displaystyle}
\newcommand{\FF}{{\cal F}}\newcommand{\AAA}{{\cal A}}
\newcommand{\CC}{{\cal C}}
\def\bra#1{\langle #1 |}
\def\ket#1{| #1\rangle }
%\vspace*{4cm}
\begin{center}
\hfill Firenze Preprint - DFF - 276/03/1997
\end{center}
\begin{center}
  \begin{Large}
  \begin{bf}
INTEGRATING A GENERIC ALGEBRA\footnote{Invited talk at the
International Seminar dedicated to the memory of D.V. Volkov, held in
Kharkov, January 5-7, 1977}\\
  \end{bf}
  \end{Large}
\end{center}
  \vspace{5mm}
\begin{center}
  \begin{large}
R. Casalbuoni\\
  \end{large}
Dipartimento di Fisica, Universita' di Firenze\\
I.N.F.N., Sezione di Firenze\\e-mail:
CASALBUONI@FI.INFN.IT
\end{center}
  \vspace{1cm}
%\begin{center}
%University of Florence - DFF XXX/12/96
%\end{center}
%\vspace{6cm}
%\begin{center}
%Firenze Preprint - DFF - 270/02/1997
%\end{center}
%$^*$ Partially supported by the Swiss National Foundation
%\newpage
\thispagestyle{empty}

%\newpage
%\begin{large}
During the last  years there has been a lot of interest in generalized
classical theories. The most typical examples are theories involving
Grassmann variables \cite{berezin}, \cite{casalbuoni} (this last paper
was largely inspired by the work in \cite{volkov}). The corresponding
path-integral quantization requires the notion of integration over the
phase-space variables. This procedure is very well known for the
particular case mentioned above \cite{berezin2}. The problem of
defining the path-integral in the general case is too much complicated
and we have limited ourselves to the first necessary step, that is to
define an integration procedure over an arbitrary algebra. This
approach  is described more completely in paper \cite{integrale}, to
which we refer for all the details. Here we will outline only the most
important steps. We want to define the integral as a linear mapping
between the given algebra and the real numbers, but we need to specify
further the properties of such a mapping. We do this by requiring the
physical principle of the combination law for the probability
amplitudes. In ordinary quantum mechanics this is mathematically
expressed through the completeness of the eigenstates of the position
operator. In order to extend this idea to the general case we use the
same approach followed in the study of non-commutative geometry
\cite{connes} and of quantum groups \cite{drinfeld}. The approach
starts from the observation that in the normal case one can
reconstruct a space from the algebra of its functions . Giving this
fact, one lifts all the necessary properties in the function space. In
this way one is able to deal with cases in which no concrete
realization of the space itself exists.

In order to see how we can lift up the completeness from the base
space to the space of functions, let us suppose that this admits  an
orthonormal set of functions.  Then, any function on the base space
can be expanded in terms of the complete set $\{\psi_n(x)\}$. It turns
out convenient to  define a generalized Fick space, $\FF$,and the
following special vector in it
\be
|\psi\rangle=\left(\matrix{\psi_0(x)\cr\psi_1(x)\cr\cdots\cr\psi_n(x)
\cr\cdots\cr}\right)
\ee
 Then, a function $ f(x)=\sum_n a_n\psi_n(x)$
 can be represented as
$ f(x)=\langle a|\psi\rangle $ where $
\langle a|=\left(a_0,a_1,\cdots,a_n,\cdots\right)
$. To write the orthogonality relation in terms of this
new formalism it is convenient to realize the complex
conjugation as a linear operation on $\FF$. In fact,
$\psi_n^*(x)$ itself can be expanded in terms of
$\psi_n(x)$, $\psi_n^*(x)=\sum_n\psi_m(x)C_{mn}$ or
$|\psi^*\rangle=C^T|\psi\rangle$. Defining a bra in
$\FF$ as the transposed of the ket $|\psi\rangle$
\be
\langle\psi|=(\psi_0(x),\psi_1(x),\cdots(x),\psi_n(x),\cdots)
\ee
the orthogonality relation becomes
\be
\int|\psi\rangle\langle\psi^*|\;dx=\int |\psi\rangle\langle\psi|C
\;dx=1
\label{newcompleteness}
\ee
Another important observation is that the orthonormal
functions define an algebra. In fact we can expand the
product of two eigenfunctions in terms of the
eigenfunctions
\be
\psi_m(x)\psi_n(x)=\sum_p c_{nmp}\psi_p(x)
\label{algebra}
\ee
with
\be
c_{nmp}=\int\psi_n(x)\psi_m(x)\psi_p^*(x)\;dx
\label{cnmp}
\ee

The relation (\ref{newcompleteness}) makes reference
only to the elements of the algebra of functions that
we have organized in the space $\FF$, and it is the key
element in order to define the integration rules on the
algebra. In fact, we can now use the algebra product to
reduce the  expression (\ref{newcompleteness}) to a
linear form. If the resulting expression has a solution
for $\int\psi_p(x)\; dx$, then we are able to define
the integration over all the algebra of functions,  by
linearity. Notice that a solution always exists, if the
constant function is in the set $\{\psi_n(x)\}$.

 The procedure we have outlined here is the one  that we will
generalize  to arbitrary algebras. Before doing that we
will consider the possibility of a further
generalization. In the usual path-integral formalism
sometimes one makes use of the coherent states  instead
of the position  operator eigenstates. In this case the
basis in which one considers the wave functions is a
basis of eigenfunctions of a non-hermitian operator $
\psi(z)=\bra\psi z\rangle$ with $ a\ket z=\ket z z$.
The wave functions of this type close an algebra, as $\langle
z^*|\psi\rangle$ do. But this time the two types of eigenfunctions are
not connected by any linear operation. In fact, the completeness
relation is defined on an algebra which is the direct product of the
two algebras
\be
\int\frac{dz^*dz}{2\pi i}\exp(-z^*z)|z\rangle\langle z^*|=1
\ee
Therefore, in similar situations, we will not define
the integration over the original algebra, but rather
on the algebra obtained by the tensor product of the
algebra times a copy. The copy corresponds to the
complex conjugated functions of the previous example.

Let us start with a generic algebra $\AAA$ with $n+1$
elements $x_i$, with $i=0,1,\cdots n$.  We do this for
simplicity, but there are no problems in letting
$n\to\infty$, or in taking a continuous index. We
assume the multiplication rules
\be
x_i x_j=f_{ijk}x_k
\ee
with the usual convention of sum over the repeated indices.
The structure constants $f_{ijk}$ define uniquely the
algebraic structure. Consider for instance the case of an
abelian algebra. In this case
\be
x_i x_j=x_j x_i\longrightarrow f_{ijk}=f_{jik}
\label{commutativity}
\ee
Or,   for  an associative algebra, from $x_i(x_j
x_k)=(x_i x_j)x_k$, one gets
\be
f_{ilm}f_{jkl}=f_{ijl}f_{lkm}
\label{associativity}
\ee

We introduce  now the space $\FF$, and the special vector
\be
| x\rangle=\left(\matrix{x_0\cr x_1\cr\cr \cdot\cr\cdot\cr
x_n\cr}\right),~~~~~~\ket x\in\FF
\ee
In order to be able to generalize properly the
discussion made for the functions, it will be of
fundamental importance to
 look for linear operators having
the vector $|x\rangle$ as eigenvector and the algebra
elements $x_i$ as eigenvalues.  This notion is strictly
related to the mathematical concept of {\bf right and
left multiplication algebras} associated to a given
algebra \cite{integrale}. The linear operators we are
looking for are defined by the relation
\be
X_i |x\rangle=|x\rangle x_i
\label{eigenvalues}
\ee
that is
\be
(X_i)_{jk}x_k=x_jx_i=f_{jik}x_k
\ee
or
\be
(X_i)_{jk}=f_{jik}
\ee
 In a complete analogous way
we can consider a bra $\bra{\tilde x}$,  defined as the
transposed of the ket $\ket x$ and we define left
multiplication through the equation
\be
\langle\tilde x|\Pi_i=x_i\langle\tilde x|
\label{lefteigenvalues}
\ee
implying
\be
(\Pi_i)_{kj}=f_{ijk}
\ee
The two matrices $X_i$ and $\Pi_i$ corresponding to
right and left multiplication are generally different.
For instance, consider  the abelian case. It follows
from eq. (\ref{commutativity})
\be
X_i=\Pi_i^T
\ee
If the algebra is associative, then from eq.
(\ref{associativity}) the following three relations can
be shown to be equivalent:
\be
X_iX_j=f_{ijk}X_k,~~~\Pi_i\Pi_j=f_{ijk}\Pi_k,~~~[X_i,\Pi_j^T]=0
\ee
The first two say that  $X_i$ and  $\Pi_i$ are linear
representations of the algebra. The third that the right and
left multiplication commute for associative algebras.

Recalling the discussion made for the functions we
would like first consider the case of a basis
originating from hermitian operators. Notice that the
generators $x_i$ play here the role of generalized
dynamical variables. It is then natural to look for the
case in which the operators $X_i$ admit both eigenkets
and eigenbras. This will be the case if
\be
\Pi_i=CX_iC^{-1}
\label{matrice_c}
\ee
that is if $\Pi_i$ and $X_i$ are connected by a
non-singular $C$ matrix. This matrix is exactly the
analogue of the matrix $C$ defined in the case of
functions.
 From eq. (\ref{lefteigenvalues}), we get
\be
\langle\tilde x|CX_iC^{-1}=x_i\langle \tilde x|
\ee
By putting
\be
\langle x|=\langle\tilde x|C
\label{conjugation}
\ee
we have
\be
\langle x|X_i=x_i\langle x|
\label{bra}
\ee
In this case, the equations (\ref{eigenvalues}) and
(\ref{bra}) show that $X_i$ is the analogue of an hermitian
operator. We will define now the
integration over the algebra by requiring that
\be
\int_{(x)}|x\rangle\langle x|=1
\ee
where $1$ is the identity matrix on the
$(n+1)\times(n+1)$ dimensional linear space of the
linear mappings on the algebra. In more explicit terms
we get
\be
 \int_{(x)}x_i  x_j=\int_{(x)}f_{ijk}x_k=(C^{-1})_{ij}
\label{integration1}
\ee
If we can invert this relation in terms of
$\int_{(x)}x_i$, we can say to have defined the
integration over the algebra, because we can extend the
operation by linearity. In particular, if ${\cal A}$ is
an algebra with identity, let us say $x_0=1$, then, by
using (\ref{integration1}), we get
\be
\int_{(x)} x_i=(C^{-1})_{0i}=(C^{-1})_{i0}
\ee
and it is always possible to define the integral.

We will discuss now the transformation properties of
the integration measure with respect to an automorphism
of the algebra. In particular, we will restrict our
analysis to the case of a simple algebra (that is an
algebra having as ideals only the algebra itself and
the null element). Let us consider an invertible linear
transformation on the basis of the algebra leaving
invariant the multiplication rules (that is an
automorphism) $ x_i'=S_{ij}x_j$ with
$x_i'x_j'=f_{ijk}x_k'$. For a simple algebra, one can
show that \cite{integrale}
\be
C^{-1}S^T C=kS^{-1}
\ee
where $k$ is a constant. It follows that the measure transforms as
\be
 \int_{(x')} =\frac 1 k  \int_{(x)}
 \label{change}
\ee
Let us consider now the case in which the automorphism
$S$ can be exponentiated in the form $S=\exp(\alpha
D)$. Then $D$ is a derivation of the algebra. If it
happens that for this particular automorphism $S$, one
has $k=1$, the integration measure is invariant, and
the integral satisfies
\be
\int_{(x)}D(f(x))=0
\label{byparts}
\ee
for any function $f(x)$ on the algebra. On the
contrary,  a derivation always defines an automorphism
of the algebra by exponentiation. So, if the
corresponding $k$ is equal to one, the equation
(\ref{byparts}) is always valid.

Of course it may happen that the  $C$ matrix does not
exist. This would correspond to the case of
non-hermitian operators as discussed  before. So we
look for a copy $\AAA^*$ of the algebra. By calling
$x^*$ the elements of $\AAA^*$, the corresponding
generators will satisfy $ x_i^*x_j^*=f_{ijk}x_k^*$. It
follows
\be
\langle{\tilde x}^*|\Pi_i=x_i^*\langle{\tilde x}^*|
\ee
Then, we define the integration rules on the tensor product
of $\AAA$ and $\AAA^*$ in such a way that the
completeness relation holds
\be
\int_{(x,x^*)}|x\rangle\langle{\tilde x}^*|=1
\label{integration2}
\ee
This second type of integration is invariant under
orthogonal transformation or unitary transformations,
according to the way in which the $^*$ operation acts
on the transformation matrix $S$. If $^*$ acts on
complex numbers as the ordinary conjugation, then we
have invariance under unitary transformations,
otherwise if $^*$ leaves complex numbers invariant,
then the invariance is under orthogonal
transformations. Notice that the invariance property
does not depend on $S$ being an automorphism of the
original algebra or not.

The two cases considered here are not mutually
exclusive. In fact, there are situations that can be
analyzed from both points of view  \cite{integrale}. We
want also to emphasize that this approach does not
pretend to be complete and that we are not going to
give any theorem about the classification of the
algebras with respect to the integration. What we are
giving is rather a set of rules that one can try to
apply in order to define an integration over an
algebra. As argued before, there are algebras that do
not admit the integration as we have defined in
(\ref{integration1}). Consider, for instance, a simple
Lie algebra. In this case we have the relation
$f_{ijk}=f_{jki}$ which implies $X_i=\Pi_i$ or $C=1$.
Then the eq. (\ref{integration1}) requires
\be
\delta_{ij}=\int_{(x)}x_ix_j=\int_{(x)}f_{ijk}x_k
\ee
which cannot be satisfied due to the antisymmetry of
the structure constants. Therefore, we can say that,
according to our integration rules, there are algebras
with a complete set of states and algebras which are
not complete. On the contrary there are many examples
in which our rules allow the definition of an
integration. We recall here, bosonic and fermionic
integration, the $q$-oscillator and the paraGrassmann
cases, and finally the integration over the algebras of
quaternions and octonions (all  these examples are
discussed in \cite{integrale}).

The work presented here is only a first approach  to
the problem of quantizing a theory defined on a
configuration space made up of non-commuting variables,
the simplest example being the case of supersymmetry.
 In order to build up the functional
integral, a second step would be necessary. In fact,
one needs a different copy of the given algebra to each
different time along the path-integration. This should
be done by taking convenient tensor products of copies
of the algebra. Given this limitation, we think,
however, that the step realized in this work is  a
necessary one in order to solve the problem of
quantizing the general theories discussed here.

\end{document}